\documentclass{aa}
\usepackage{graphics}
\usepackage{psfig}
\begin{document}

\newcommand{\eff}{\mbox{Effelsberg}}
\newcommand{\gb}{\mbox{Green Bank}}
\newcommand{\nan}{\mbox{Nan\c{c}ay}}
\newcommand{\as}[2]{$#1''\,\hspace{-1.7mm}.\hspace{.1mm}#2$}
\newcommand{\am}[2]{$#1'\,\hspace{-1.7mm}.\hspace{.0mm}#2$}
\def\approxlt{\lower.2em\hbox{$\buildrel < \over \sim$}}
\def\approxgt{\lower.2em\hbox{$\buildrel > \over \sim$}}
\newcommand{\dgr}{\mbox{$^\circ$}}   
\newcommand{\E}[1]{\mbox{${}\,10^{#1}{}$}}
\newcommand{\etal}{et al. }
\def \eg{{e.g.}}
\newcommand{\grd}[2]{\mbox{#1\fdg #2}}
\newcommand{\gsim}{\stackrel{>}{_{\sim}}}
\newcommand{\lsim}{\stackrel{<}{_{\sim}}}
\newcommand{\Ha}{\mbox{H$\alpha$}}
\newcommand{\HI}{\mbox{H\,{\sc i}}}
\newcommand{\HIbf}{\mbox{H\hspace{0.155 em}{\footnotesize \bf I}}}
\newcommand{\HIit}{\mbox{H\hspace{0.155 em}{\footnotesize \it I}}}
\newcommand{\HIsl}{\mbox{H\hspace{0.155 em}{\footnotesize \sl I}}}
\newcommand{\HIss}{\mbox{H\,{\sc i}}}
\newcommand{\HII}{\mbox{H\,{\sc ii}}}
\def \ie{{i.e.}}
\newcommand{\kms}{\mbox{ km\,s$^{-1}$}}
\newcommand{\kmsMpc}{\mbox{ km\,s$^{-1}$\,Mpc$^{-1}$}}
\newcommand{\LB}{\mbox{$L_{B}$}}
\newcommand{\LBnul}{\mbox{$L_{B}^0$}}
\newcommand{\LBsun}{\mbox{$L_{\odot,B}$}}
\newcommand{\Lsun}{\mbox{$L_{\odot}$}}
\newcommand{\LsunMsun}{\mbox{$L_{\odot}$/${M}_{\odot}$}}
\newcommand{\LFIR}{\mbox{$L_{FIR}$}}
\newcommand{\LFIRLB}{\mbox{$L_{FIR}$/$L_{B}$}}
\newcommand{\LFIRLBnul}{\mbox{$L_{FIR}$/$L_{B}^0$}}
\newcommand{\LFIRLsun}{\mbox{$L_{FIR}$/$L_{\odot,Bol}$}}
\def\lir{{\hbox {$L_{IR}$}}}
\def\lco{{\hbox {$L_{CO}$}}}
\def \ls{\hbox{$L_{\odot}$}}
\def \ms{\hbox{$M_{\odot}$}}
\newcommand{\MHI}{\mbox{${M}_{HI}$}}
\newcommand{\MHILB}{\mbox{${M}_{HI}$/$L_{B}$}}
\newcommand{\MHILBnul}{\mbox{${M}_{HI}$/$L_{B}^0$}}
\newcommand{\Msun}{\mbox{${M}_\odot$}}
\newcommand{\MsunLsun}{\mbox{${M}_{\odot}$/$L_{\odot,Bol}$}}
\newcommand{\MsunLBsun}{\mbox{${M}_{\odot}$/$L_{\odot,B}$}}
\newcommand{\MT}{\mbox{${M}_{ T}$}}
\newcommand{\MTLBnul}{\mbox{${M}_{T}$/$L_{B}^0$}}
\newcommand{\MTLBsun}{\mbox{${M}_{T}$/$L_{\odot,B}$}}
\newcommand{\mi}{\mbox{$\mu$m}}
\newcommand{\NH}{\mbox{N$_{HI}$}}
\newcommand{\OIII}{\mbox{[O\,{\sc iii}]}}
\newcommand{\s}{\mbox{$\sigma$}}
\newcommand{\Tb}{\mbox{$T_{b}$}}
\newcommand{\tis}[2]{$#1^{s}\,\hspace{-1.7mm}.\hspace{.1mm}#2$}
\newcommand{\vhel}{\mbox{$V_{hel}$}}
\newcommand{\vrot}{\mbox{$v_{rot}$}}
\def\la{\mathrel{\hbox{\rlap{\hbox{\lower4pt\hbox{$\sim$}}}\hbox{$<$}}}}
\def\ga{\mathrel{\hbox{\rlap{\hbox{\lower4pt\hbox{$\sim$}}}\hbox{$>$}}}}

\thesaurus{03(11.04.1;   % Galaxies: distances and redshifts
              11.07.1;   % Galaxies: general
              11.09.4;   % Galaxies: ISM
              13.19.1)}  % Radio lines: galaxies

\title{H{\large \bf I} line observations of luminous infrared galaxy mergers}

\author{W. van Driel\,\inst{1,2}, Yu Gao\,\inst{3},
        D. Monnier-Ragaigne\,\inst{1,2} }

\offprints{W. van Driel, e-mail: wim.vandriel@obspm.fr
}

\institute{Unit\'e Scientifique \nan, USR CNRS B704,
 Observatoire de Paris, 5 place Jules Janssen, F-92195 Meudon, France
\and
DAEC, UMR CNRS 8631, Observatoire de Paris, 5 place Jules Janssen,
 F-92195 Meudon, France
\and
IPAC, Caltech, MS 100-22, 770 South Wilson Ave., Pasadena, CA 91125, U.S.A.
}

\date{\it Received 22/9/2000; accepted 22/12/2000}
\maketitle

\markboth{{W. van Driel et al.: \HI\ observations of luminous infrared 
           galaxy mergers}}{}

\begin{abstract}
A total of 19 luminous infrared galaxy mergers, with $L_{\rm IR} \approxgt\
2\times10^{11}~\ls$, for $H_0=75$~\kms~Mpc$^{-1}$, have been observed in the 
\HI\ line at \nan\ and four of them were observed at Arecibo as well. 
Of these 19, ten had not been observed before. Six were clearly 
detected, one of which for the first time. The objective was to statistically  
sample the \HI\ gas mass in luminous infrared mergers along a 
starburst merger sequence 
where the molecular CO gas content is already known. We also searched the 
literature for \HI\ data and compared these with our observations.
\end{abstract}

\section{Introduction}  % 1
Most luminous infrared galaxies (LIGs), with infrared luminosities
$L_{\rm IR} \approxgt 2\times10^{11}~\ls$, for $H_0=75$~\kms~Mpc$^{-1}$, 
are gas-rich, closely interacting/merging systems (\eg, van den Broek et al. 1991; 
Sanders 1992; Murphy et al. 1996)  with high star formation rates and efficiencies.
They represent the dominant population among objects with
$L_{\rm bol} \approxgt 2\times 10^{11} \ \ls$ in the local
universe (Sanders \& Mirabel 1996). 

Most of the LIGs' luminosity is radiated by dust in the far-infrared.
Evidence in favor of a starburst origin for the high IR luminosity
is clearly mounting (\eg, from molecular gas content, Downes \& Solomon 
1998; Solomon et al. 1997; radio continuum imaging, Condon et al. 
1991; Smith et al. 1998; ISO mid-IR spectroscopy, Genzel et al. 1998; 
and ground based mid-IR and  HST/NICMOS near-IR imaging,
Soifer et al. 2000; Scoville et al. 2000).
Although an AGN might also contribute to the high IR 
luminosity of a small fraction LIGs (\eg, Sanders et al. 1988; 
Veilleux et al. 1997; 1999; Sanders 1999), the observed abundant supply 
of {\it dense} molecular gas traced by high dipole moment molecules 
like HCN, usually found only in star formation cores (Solomon et al.
1992; Gao 1996; Gao \& Solomon 2000a, b) indicates that almost all 
LIGs are ideal stellar nurseries.  

A clear correlation between the CO(1-0)
luminosity (a measure of molecular gas mass M(H$_2$)) and the projected
separation of merger nuclei (an indicator of merging stages) 
has been shown for a sample of 50 LIG mergers 
(Gao \& Solomon 1999), indicating that 
the molecular gas content decreases as merging advances. 
This suggests that the molecular
gas content is being rapidly depleted due to the starbursts as merging
progresses. This was the first evidence connecting the depletion of 
molecular gas with starbursts in interacting galaxies. 
The question arises if a similar relationship exists between the 
\HI\ gas mass and the merger interaction stages.

Only loose correlations were found between the \HI\ gas mass
and \lir\ in previous \HI\ line studies of LIGs samples of LIGs
(Garwood et al. 1987; Jackson et al. 1989; Martin et al. 1991;
Mirabel \& Sanders 1988; Young et al. 1986),
unlike the tight correlation found between
\lir\ and the CO gas mass. A tighter correlation was
found (Kennicutt 1998) between the \LFIR/area and \MHI/area, however.
This may imply that the 
more diffuse \HI\ gas is being depleted as well, like the molecular
gas component, as merging advances.

The M(H$_2$)/M(\HI) ratio also increases with \lir\ for LIGs 
(Mirabel \& Sanders 1989). However, whether this ratio
increases or not as the merging progresses through different stages 
has not yet been explored. With new \HI\ data 
accumulated for a statistically significant sample of LIGs
arranged along a starburst merger sequence (e.g., by the degree
of nuclear separation and the merging morphology compared with
numerical simulations for mergers), we can ultimately examine how the  
M(H$_2$)/M(\HI) ratio changes as the merger-induced 
starburst proceeds in mergers of different stages, 
since the molecular gas contents of our sample
galaxies are already known. 

Just over a dozen galaxies in the Gao \& Solomon (1999) CO sample of 
LIG mergers have 21cm \HI\ line measurements reported in the literature. 
In order to enlarge the sample available for a statistical study of the 
relationship  between the \HI\ gas mass and the evolution of the merging 
process as the starburst proceeds along a starburst merger sequence, 
we have conducted new \HI\ observations. 
Here, we present the results of new \nan\ and Arecibo 21cm observations 
of 19 LIGs in the sample (for the sample selection, see Sect. 2).

We will explore possible correlations with the global \HI\ properties 
of LIGs in detail in another paper (Gao et al. in preparation). For this, 
we will use all \HI\ observations available for LIGs in the 
CO survey of Gao \& Solomon (1999), including 
spectra from an unpublished Arecibo \HI\ survey of LIGs 
(I.F. Mirabel 1997, private communication).

% ----- Figure 1 ------------------------------------------------------
%
\begin{figure*}
\vspace*{0.25cm}
% \addtocounter{figure}{-1}
\psfig{figure=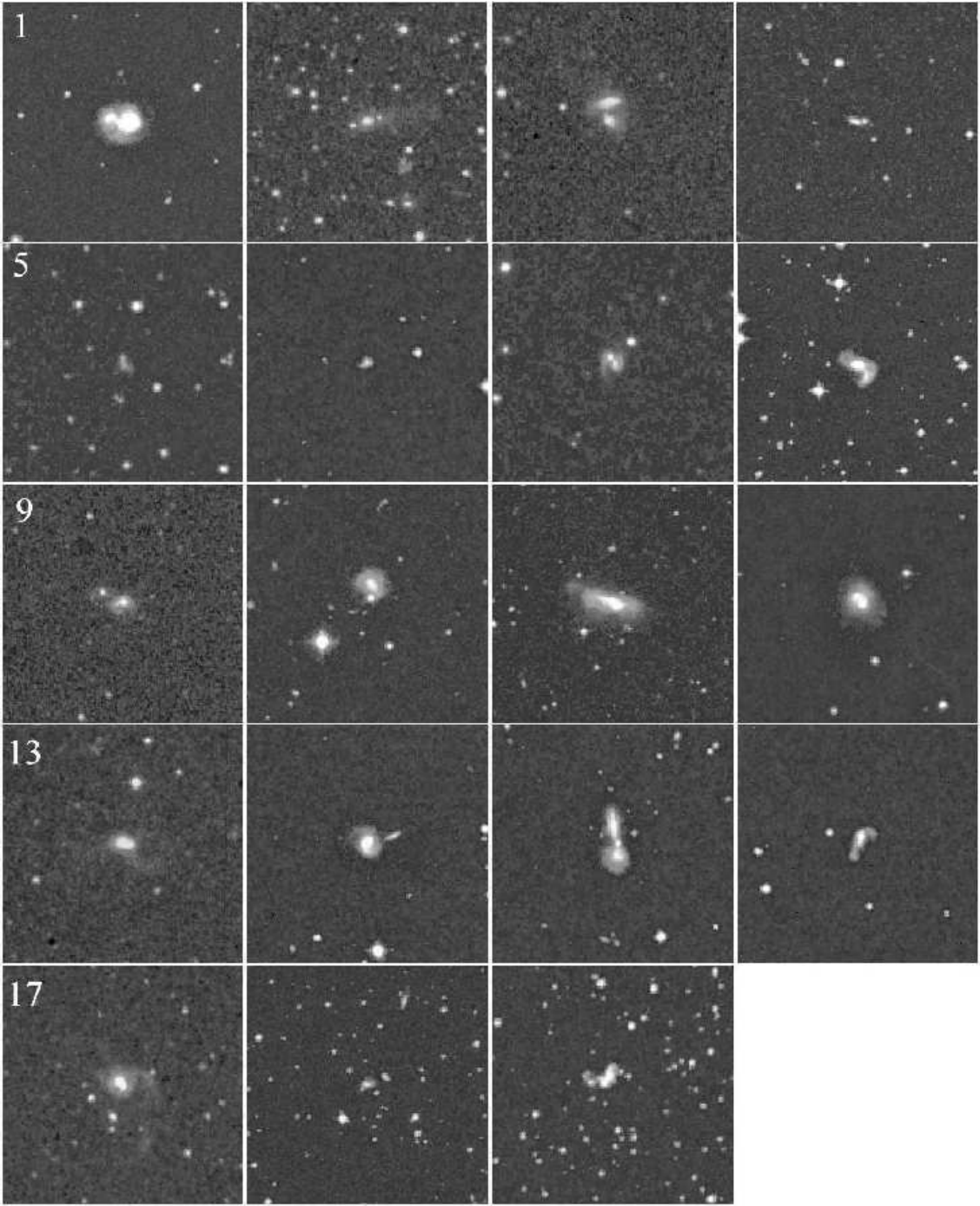,width=15cm}
\caption [] {Mosaic of Digital Sky Survey images of the 19 infrared luminous
merger galaxies observed. The dimensions of the individual images are 5$'\times$5$'$.
The order of the galaxies is the same as that used in the Tables.
}
\end{figure*}

\section{Sample selection -- luminous infrared galaxies along a 
 starburst merger sequence}  % 2
Although there have been extensive \HI\ and CO observations of LIGs,
including synthesis imaging (\eg, Mirabel \& Sanders 1988; Martin et al. 1991;
Scoville et al. 1991; Gao 1996; Gao et al. 1997; Lo, Gao \& Gruendl 1997; 
Downes \& Solomon 1998; Hibbard \& Yun 1996, 1999a, 1999b; Gao \& Solomon 1999;
Gao et al. 1999, 2000a, b), there has been no 
systematic attempt to trace the gas properties (both atomic and molecular) 
during the galaxy-galaxy merger sequence starting with 
systems roughly separated by a galactic diameter; e.g., just 4 galaxies in the 
``Toomre Sequence'' (Toomre \& Toomre 1972; Toomre 1977), representing a merger 
sequence, were observed by Hibbard \& van Gorkom (1996). 

However, observations of only the atomic gas along the ``Toomre Sequence'' 
is not sufficient to trace the starburst processes occurring during a 
galaxy-galaxy merger, as stars are forming in molecular clouds which are 
probably embedded in the more diffuse atomic hydrogen gas, and \HI\ itself 
is not a tracer of the star forming gas. Moreover, the galaxy mergers in the 
``Toomre Sequence'' are optically selected and have 
large variations in their gas content and do not easily fit into a sequence
of gas conversion into stars. 

Although the ``Toomre Sequence''
nicely reflects the dynamical transformation of two spiral galaxies into 
a merged elliptical, no ultraluminous starburst is
included in this sequence -- the four most infrared luminous objects
on the sequence are NGC 6621/2, NGC 7592, NGC 2623 and NGC 3256, with,
respectively, log(\lir/\ls)=11.26, 11.33, 11.55 and 11.57.
Therefore, although the triggering of a starburst by the
merging of two spiral galaxies is probably represented by the 
``Toomre Sequence'', it poorly samples the ultraluminous starburst
phase of gas-rich mergers along the merger sequence. 
In short, the ``Toomre Sequence'' 
is most likely a morphological sequence, 
and it tells us little about the conversion of
gas into stars during mergers.
Indeed, there are many close mergers where the progenitors are not 
gas-rich galaxies and these may never become LIGs 
(e.g., Bushouse et al. 1988; Kennicutt et al. 1987).

% -----------------------------------------------------------------
%
% --- Table 1: Basic data
%
\begin{table*}
\bigskip
{\footnotesize
\begin{tabular}{lllrrrlllll}
\multicolumn{11}{l}{{\bf Table 1.} Basic optical data for the sample galaxies.}\\
\smallskip \\
\hline
\vspace{-2 mm} \\
No. & Ident & R.A. & Dec & $V_{opt}$ & err & $B_T$ & D$\times$d & Morph & Spect & sep \\   
 &  & \multicolumn{2}{c}{(1950.0)} &  &  &  &  &  &  &  \\    
 &  & hh mm ss.s & dd mm ss & km/s & km/s & mag & $'\,\times \,'$ &  &  &  $''$ \\ 
\vspace{-2 mm} \\
\hline
\vspace{-2 mm} \\
%                            NED          NED       LEDA            LEDA        LEDA        NED       NED
1  & IC 1623A          & 01 05 19.9 & -17 46 29 &   6042  &  72  &  13.70  &  1.0\,\,0.8  & Sp:     & \HII\ & 16.0 \\ % HI
   & IC 1623B          & 01 05 22.0 & -17 46 18 &   5840  &  60  &  15.00  &  0.8\,\,0.6  &         & \HII\ &      \\
2  & IRAS 02483+4302   & 02 48 19.8 &  43 02 53 &  15475  & 105  &  16.75  &  0.6\,\,0.3  &         &       &  3.8 \\
3  & UGC 2369A         & 02 51 15.9 &  14 46 24 &   9463  & 152  &  15.39  &  0.7\,\,0.3  & Sb      &       & 22.5 \\ 
   & UGC 2369B         & 02 51 15.9 &  14 46 03 &   9380  & 118  &  15.54  &  0.7\,\,0.4  &         &       &      \\
4  & IRAS 03359+1523   & 03 35 57.5 &  15 23 10 &  10626  &  45  &  16.58  &  0.4\,\,0.2  &         & \HII\ & 10.0 \\
5  & IRAS 04232+1436   & 04 23 15.2 &  14 36 53 &  23972  &  60  &  18.60  &  0.3\,\,0.2  &         &       &  4.6 \\  
6  & IRAS 08572+3915   & 08 57 12.9 &  39 15 39 &  17480  &  42  & (14.92) &              &        & L:/Sy2 &  5.5 \\ 
7  & IRAS 10035+4852   & 10 03 35.7 &  48 52 23 &  19396  &  50  &  16.13  &  0.6\,\,0.4  &         &       &  8.7 \\
8  & IC 2545           & 10 03 53.0 & -33 38 30 &  10245  &  47  &  15.25  &  0.65\,\,0.4 &         &       &  5.0 \\ % HI
   &                   & 10 03 52.0 & -33 38 42 &         &      & (16.83) &              &         &       &      \\
9  & IRAS 10565+2448W  & 10 56 35.5 &  24 48 43 & (13160) & (81) &         &              &         &       &  8.0 \\
   & IRAS 10565+2448NE & 10 56 35.5 &  24 48 43 & (12937) & (37) & (16.0)  &              &         &       &      \\
10 & Mrk 238           & 12 59 20.7 &  65 16 06 &  14816  & 171  &  15.92  &  0.8\,\,0.7  &         &       & 14.8 \\
11 & IRAS 13001-2339   & 13 00 10.5 & -23 39 13 &   6417  & 130  &  14.68  &  1.3\,\,0.5  &  I0:p   &       &  6.5 \\
12 & NGC 5256A         & 13 36 14.5 &  48 31 47 &   8316  &  55  & (14.1)  &  0.5\,\,0.4  &  Comp.p & Sy2   &  9.5 \\
   & NGC 5256B         & 13 36 15.0 &  48 31 54 &   8419  &  48  &  13.67  &  0.7\,\,0.6  &  Comp.p & L     &      \\
13 & Mrk 463           & 13 53 39.6 &  18 36 57 & (15289) &      & (17.75) &              &         & Sy2   &  3.9 \\
   &                   & 13 53 39.8 &  18 36 58 & (14990) &      & (17.25) &              &         & Sy1   &      \\
14 & IC 4395           & 14 15 06.2 &  27 05 18 &  10947  &   7  &  14.84  &  1.1\,\,0.8  &  S?     & \HII\ &  6.0 \\ % HI
15 & UGC 9618A         & 14 54 47.9 &  24 48 28 &  10029  &  10  &  15.30  &  0.8\,\,0.7  &  Sc     &       & 41.6 \\ % HI
   & UGC 9618B         & 14 54 48.2 &  24 49 03 &  10157  &  53  &  15.70  &  0.8\,\,0.3  &  Sb     & \HII/L &     \\
16 & Mrk 848           & 15 16 19.5 &  42 55 30 &  12049  & 131  &  15.95  &  0.7\,\,0.3  &         &       &  6.5 \\
17 & NGC 6090          & 16 10 24.1 &  52 35 00 &  (9062) & (30) & (15.0)  & (0.3\,\,0.3) &  Sd p   & \HII\ &  6.5 \\ % HI
   &                   & 16 10 24.6 &  52 35 04 &  (8822) & (30) & (14.5)  & (0.4\,\,0.2) &  Sd p   & \HII\ &      \\
18 & IRAS 20010-2352   & 20 01 04.2 & -23 52 25 &  15194  &  69  &  17.07  & 0.4\,\,0.3   &         & Sy2   &  8.9 \\  
19 & II Zw 96          & 20 55 04.6 &  16 56 07 & (10630) & (81) & (16.41) &              &         & \HII\ & 11.0 \\ % HI
   &                   & 20 55 05.0 &  16 55 58 & (10770) & (81) & (15.20) &              &         &       &      \\
\vspace{-2 mm} \\
\hline
\vspace{-2 mm} \\
\multicolumn{11}{l}{Note: the values in brackets are from the NED database, all
others are weighted means from the LEDA database.} \\
\end{tabular}
}
\normalsize
\end{table*}
% -----------------------------------------------------------------

The ideal sample to trace a merger sequence leading to the formation
of an ultraluminous starburst would contain gas-rich galaxies of 
various merging stages that initially started 
with roughly comparably, large gas contents. 
Since this is impossible to ascertain, we selected LIGs where 
the 2 progenitors can be identified from CCD images. 
By measuring the \HI\ content as a function of the
merger separation for a sample with known molecular gas
properties, the time dependence of the total (\HI+CO) gas 
properties can then be traced statistically in a starburst merger sequence.

50 LIGs with \ $L_{IR}~\approxgt~ 2\times 10^{11}~\ls, \ 
2'' \approxlt \ S_{sep} \approxlt (D_1+D_2)/2$ \ (where $S_{sep}$ is the
separation between nuclei and $D_1$ and $D_2$ are the major axis diameters of 
the two galaxies) have been selected for the CO study of Gao \& Solomon (1999),
and we already have molecular gas measurements for all these galaxies. 
Here, we aim at establishing a large homogeneous sample
of LIGs which have \HI\ measurements as well. 
For observation at \nan, we excluded all high redshift ($cz>20,000\kms$) 
sources, deemed {\it a priori} too weak in \HI\ for a 100m-class telescope, 
and all southern sky ($\delta<-$39$^\circ$) objects
inaccessible from the site, thus reducing the sample to 30 LIGs.
By further excluding sources around the Galactic Center's right ascension, 
where the transit telescope is too heavily oversubscribed with
high-priority targets, and by avoiding the repetition of previously 
well-detected sources, we ended up with a total sample of 19 LIGs 
from our CO sample (see Table 1). For all of these we obtained new global 
\HI\ spectra.

\section{Observations and data reduction}  % 3
\subsection{Nan\c{c}ay observations}
The \nan\ telescope is a meridian transit-type instrument of the Kraus/Ohio State 
design, consisting of a fixed spherical mirror, 300 m long and 35 m high,
a tiltable flat mirror (200$\times$40 m), and a focal carriage moving along a 
90 m long curved rail track, which allows the tracking of a source on the 
celestial equator for about 1 hour. Located in the centre of France,
it can reach declinations as low as -39$^\circ$. It has an effective collecting 
area of about 7000~m$^{2}$ (equivalent to a 94-m diameter parabolic dish). 
Due to the elongated mirror geometry, at 21-cm wavelength it has a 
half-power beam width of \am{3}{6}~E-W$\times$22$'$~N-S for declinations 
below 20$^{\circ}$, increasing to 34$'$~N-S at $\delta$=65$^{\circ}$,
the highest declination in the present survey 
(E. G\'erard, private comm.; see also Matthews \& van Driel 2000). 
Typical system temperatures were $\sim$40~K for 
our project.

% ----- Figure 2 ------------------------------------------------------
%
\begin{figure*}
\vspace*{0.5cm}
% \addtocounter{figure}{-1}
\psfig{figure=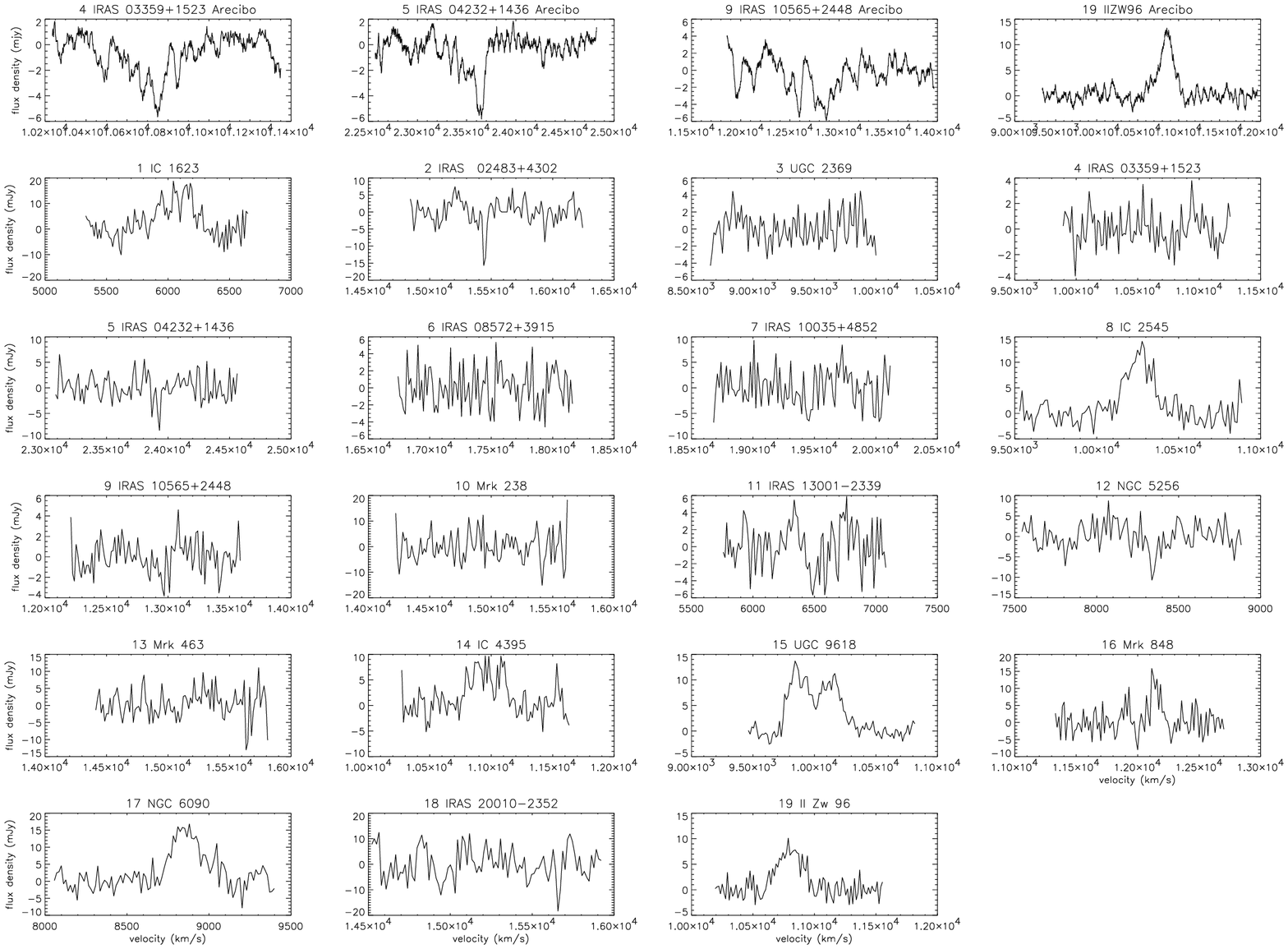,width=18cm,angle=90}
\caption [] {Arecibo (top row) and \nan\ 21-cm \HI\ line spectra.  
Velocity resolution is 27.0 \kms\
for the 4 Arecibo profiles and 15.8 \kms\ for the 19 \nan\ spectra, 
radial heliocentric velocities are according to the optical convention. 
The spectra are shown in the same order as in Fig.~1 and Table~1.
The line signal in the \nan\ Mrk 848 spectrum is caused by residual radio interference,
see Sect. 4.
}
\end{figure*}

Our observations were made throughout the period December 1998 - October 1999, 
using a total of about 150 hours of telescope time. We obtained our observations 
in total power (position-switching) mode using consecutive pairs 
of two-minute on- and two-minute off-source integrations. Off-source integrations 
were taken at approximately 2$^m$~E of the target position.
The autocorrelator was divided into two pairs
of cross-polarized receiver banks, each with 512 channels and a 6.4~MHz 
bandpass. This yielded a channel spacing of 2.64~\kms\ and an effective
velocity resolution of $\sim$3.3~\kms, which was smoothed
to a channel separation of 13.2 and a velocity resolution of 15.8 \kms\
during the data reduction, in order to search for faint features. The centre 
frequencies of the two banks were set to the known redshifted \HI\ 
frequency of the target; all galaxies had existing optical and CO redshifts.

We reduced our \HI\ spectra using the standard DAC and SIR
spectral line reduction packages available at the \nan\ site. 
With this software we subtracted baselines (generally third order 
polynomials) and averaged the two receiver polarizations.
To convert from units of $T_{sys}$ to flux density in mJy
we used the calibration procedure described in Matthews et al. (2001), 
see also Matthews et al. (1998) and Matthews \& van Driel (2000).
This procedure yields an internal calibration accuracy of about $\pm$15\% 
near the rest frequency of the 21cm line, and $\pm$25\% 
for the highest redshift source in the sample, at V=24,000 \kms\ (1300 MHz),
as less standard calibration observations were made at these infrequently
observed high velocities.

% -----------------------------------------------------------------
%
% Table 2. HI results
% 
\begin{table*}
\bigskip
{\scriptsize
\begin{tabular}{lllrrrrrrll}
\multicolumn{11}{l}{\footnotesize {\bf Table 2.} Basic \HI\ data - new results and literature 
values} \\
\smallskip \\
\hline
\vspace{-2 mm} \\
No. & Ident & R.A. & Dec & rms & $I_{HI}$ & $V_{HI}$ & $W_{50}$ & $W_{20}$ & Tel & Ref \\  
 &  & \multicolumn{2}{c}{(1950.0)} &  &  &  &  &  &  &  \\    
 &  & hh mm ss.s & dd mm ss & mJy & Jy\kms & \kms & \kms & \kms &  &  \\ 
\vspace{-2 mm} \\
\hline
\vspace{-2 mm} \\
1  & IC 1623         &  01 05 21.0 &  -17 46 23 &  5.51 &    4.56 &  6053$\pm$20 & 287 & 331 & N & * \\
   &                 &             &            &       &    3.30 &  6078 & 244 & 380 & N & Ma91 \\ 
2  & IRAS 02483+4302 &  02 48 19.8 &   43 02 53 &  4.90 & $<$2.45 &       &     &     & N & *    \\
3  & UGC 2369        &  02 51 15.8 &   14 46 14 &  2.43 & $<$2.19 &       &     &     & N & *    \\ 
   &                 &             &            &       &    0.48 & 9761 & 229: & --- & A & Ha97 \\  
   &                 &             &  & & {\it abs.} & {\it 9406} & --- & {\it 274} & A & MS88 \\
4  & IRAS 03359+1523 &  03 35 57.9 &   15 23 11 &  1.76 & $<$0.83 &       &     &     & N & *    \\
   &                 &             &            & 1.11 & {\it abs.} & {\it 10725}& {\it 189} & --- & A & * \\
   &                 &             & & & {\it abs.} & {\it 10717} & --- & {\it 294} & A & MS88 \\
5  & IRAS 04232+1436 &  04 23 15.2 &   14 36 53 &  3.43 & $<$2.74 &       &     &     & N & *    \\
   &                 &             &            &  0.59 & $<$0.47 &       &     &     & A & * \\
6  & IRAS 08572+3915 &  08 57 12.9 &   39 15 39 &  3.11 & $<$1.68 &       &     &     & N & *    \\
7  & IRAS 10035+4852 &  10 03 35.7 &   48 52 23 &  4.85 & $<$2.43 &       &     &     & N & *    \\    
8  & IC 2545         &  10 03 52.3 &  -33 38 24 &  3.25 &  2.56   & 10257$\pm$16 & 182 & 228 & N & * \\
9  & IRAS 10565+2448 &  10 56 36.2 &   24 48 40 &  2.13 & $<$1.28 &       &     &     & N & *    \\
   &                 &             &            & 1.17 & {\it abs} & {\it 12825:} &  --- & --- & A & * \\
   &                 &             &            &       & $>$0.11 & 13100 & --- & $>$137 & A & MS88 \\ 
   &                 &             & & & {\it abs.} & {\it 12700} & --- & {\it 646} & A & MS88 \\
10 & Mrk 238         &  12 59 20.7 &   65 16 06 &  7.74 & $<$5.42 &       &     &     & N & *    \\
11 & IRAS 13001-2339 &  13 00 10.5 &  -23 39 13 &  3.43 & $<$3.09 &       &     &     & N & *    \\
12 & NGC 5256        &  13 36 15.0 &   48 31 51 &  4.39 & $<$2.51 &       &     &     & N & *    \\
13 & Mrk 463         &  13 53 39.7 &   18 36 57 &  5.93 & $<$3.55 &       &     &     & N & *    \\ 
   &                 &             &            &       &    0.20 & 15230 & --- & 200 & A & Hu87 \\ 
   &                 &             &            &    & $\leq$0.24 &       &     &     & A & BB83 \\
   &                 &             &            &       &    6.0  & 14702 &     &     & G & He78 \\  
14 & IC 4395         &  14 15 06.2 &   27 05 18 &  3.26 &    2.54 & 10954$\pm$18 & 308 & 340 & N & * \\
   &                 &             &            &       &    1.30 & 10949 & 285 & 328 & N & Ma91 \\ 
   &                 &             &            &       &   0.62: & 10946: & 199: & --- & A & Fr88 \\   
15 & UGC 9618        &  14 54 48.3 &   24 48 42 &  1.42 &    5.43 &  9980$\pm$11 & 419 & 529 & N & * \\
   &                 &             &            &       &    3.86 & 10029 & 317 & 442 & N & Ma91 \\
   &                 &             &            &       & $>$4.66 &  9982 & --- & 513 & A & MS88 \\
   &                 &             &  & & {\it abs.} & {\it 9990} & --- & {\it 215} & A & MS88 \\
16 & Mrk 848         &  15 16 19.5 &   42 55 30 &  4.77 &     --- &       &     &     & N & *    \\
17 & NGC 6090        &  16 10 24.0 &   52 35 04 &  3.57 &    4.30 &  8901$\pm$28 & 203 & 357 & N & * \\
   &                 &             &            &       &    3.21 &  8841 & 211 & 232 & N & Ma91 \\ 
   &                 &             &            &       &    3.4  &  8871 & --- & 348 & G & HR89 \\
   &                 &             &            &       &    5.29 &  8873 & --- & --- & G & Bu87 \\
18 & IRAS 20010-2352 &  20 01 04.2 &  -23 52 25 &  8.35 & $<$9.19 &       &     &     & N & *    \\
19 & II Zw 96        &  20 55 04.8 &   16 56 02 &  2.15 &    2.26 & 10803$\pm$17 & 270 & 315 & N & * \\
   &                 &             &            &  1.16 &    2.55 & 10860$\pm$9  & 189 & 412 & A & *    \\
   &                 &             &            &       &    2.34 & 10822 & 211 & 320 & A & GH93 \\
   &                 &             &            &       & $>$2.08 & 10810 & --- & 263 & A & MS88 \\
   &                 &             & & & {\it abs.} & {\it 11132} & --- & {\it 300} & A & MS88 \\
\vspace{-2 mm} \\
\hline
\vspace{-2 mm} \\
\multicolumn{11}{l}{Notes: `:' denotes an uncertain value; values in {\it italics}
                    are for absorption line profiles; } \\
\multicolumn{11}{l}{upper limits to $I_{HI}$ are 2$\sigma$ values for flat-topped profiles
                    with linewidths assumed equal to the CO line values (see Table 3).} \\
 \\
\multicolumn{11}{l}{\HI\ references and telescope codes:}\\
 \\
% \vspace{-2 mm} \\
% \hline \\
% \vspace{-2 mm} \\
BB83  &  \multicolumn{2}{l}{Bieging \& Biermann (1983)} &
Bu87  &    \multicolumn{7}{l}{Bushouse (1987)} \\
Fr88  &  \multicolumn{2}{l}{Freudling et al. (1988)} &
GH93  &    \multicolumn{7}{l}{Giovanelli \& Haynes (1993)} \\
Ha97  &  \multicolumn{2}{l}{Haynes et al. (1997)} &
He78  &    \multicolumn{7}{l}{Heckman et al. (1978)} \\
HR89  &  \multicolumn{2}{l}{Huchtmeier \& Richter (1989)} &
Hu87  &    \multicolumn{7}{l}{Hutchings et al. (1987)} \\
Ma91  &  \multicolumn{2}{l}{Martin et al. (1991)} &
MS88  &    \multicolumn{7}{l}{Mirabel \& Sanders (1988)} \\
$\star$ &  \multicolumn{2}{l}{present paper} &
        &  \multicolumn{7}{l}{ } \\
 \\
% \vspace{-2 mm} \\
% \hline \\
% \vspace{-2 mm} \\
\multicolumn{2}{l}{ A\,\, Arecibo} &
\multicolumn{2}{l}{ N\,\, \nan} &
\multicolumn{7}{l}{ G\,\, Green Bank 90m} \\
\vspace{-2 mm} \\
\hline \\
\end{tabular}
}
\normalsize
\end{table*}
% -----------------------------------------------------------------

Another consequence of observing redshifted objects down to a frequency of
1300 MHz, far outside the protected 1400-1427 MHz frequency 
band allocated on a primary base to the Radio Astronomy Service (e.g., CRAF Handbook
on Radio Astronomy 1997), is the
occurrence of strong radio interference (RFI) signals, which were especially
noticeable in the 10,000--12,000 \kms\ range (see, e.g., the remarks on
Mrk 848 in Sect. 4.1).

\subsection{Arecibo observations}
During an observing run at Arecibo from 1 to 13 november, 2000, we obtained \HI\
spectra of the 4 sample galaxies which could be observed within the declination
range of the 305-m diameter telescope and the allocated sidereal time slots: 
IRAS 03359+1523, 04232+1436, 10565+2448 and II Zw 96. 
Data were taken with the $L$-narrow receiver,
set to linear polarisation, with two polarisations, 4096 channel subcorrelators
and nine channel sampling.
Each subcorrelator had a 12.5 MHz bandpass, resulting in a 1.3 \kms\ resolution. 
Observations were taken in 3 minute ON/OFF pairs, followed by a
10 sec ON/OFF calibration pair. 
The data were reduced using the ANALYZ software package (Deich 1990). The two
polarisations were combined, and corrections were applied
for the variations in the gain and system temperature of the telescope with zenith
angle. A 21-channel boxcar velocity smoothing was applied, resulting
in a resolution of 27.0 \kms. Baselines were then fitted to the data using the 
GALPAK package adapted by K. O'Neil, which was also used to correct the velocities 
to the heliocentric system and to determine the global \HI\ line parameters.

\section{Results and Discussion}  % 4
The 19 \nan\ and the 4 Arecibo \HI\ spectra are shown in Fig.~2.
Raw, measured radial velocities, $V_{HI}$ (in the optical convention) and their
uncertainties, integrated line fluxes, $I_{HI}$,
velocity widths at 50\% and 20\% of peak maximum, $W_{50}$ and $W_{20}$, 
and rms noise levels of our spectra were determined using the SIR
data reduction package for the \nan\ data and the adapted GALPAK package for the
Arecibo data (see Matthews et al. 2001, and references therein).

No corrections have been applied to these values for, e.g., instrumental resolution
or cosmological stretching (e.g., Matthews et al. 2001).
We estimated the uncertainties, $\sigma_{V_{HI}}$, in the central \HI\ velocities
following Fouqu\'e et al. (1990):
\begin{equation}
\sigma_{V_{HI}} = 4 R^{0.5}P_{W}^{0.5}X^{-1}\,\, [km/s]
\end{equation}
where R is the instrumental resolution in \kms, P$_{W}$=(W$_{20}$--W$_{50}$)/2
$[$km/s$]$ and X is the signal-to-noise ratio of a spectrum, which we defined as
the ratio of the peak flux density and the rms noise. According to Fouqu\'e et al.,
the uncertainty in the linewidths is 2$\sigma_{V_{HI}}$ for W$_{50}$ and
3$\sigma_{V_{HI}}$ for W$_{20}$.
These \HI\ profile parameters are listed in Table 2, together with values
found in the literature. 
For non-detections, the estimated upper limits to the integrated \HI\ fluxes are 
2$\sigma$ values for flat-topped profiles with a linewidth equal to those measured
for the CO line emission of the objects, either FWHM or half of FWZI values 
(see Table 3, for the CO data references see Gao \& Solomon 1999).
Derived global properties of the 19 sample galaxies are listed in Table 3;
the CO line luminosity values are from Gao \& Solomon (1999).

Of these 19 luminous infrared mergers, 10 were observed here for the first time 
in the 21cm line (IC 2545; IRAS 02483+4302, 04232+1436, 08572+3915, 
10035+4852, 13001-2339 and 20010-2352; Mrk 238 and 848; NGC 5256 - i.e., 
Nos. 2, 5, 6, 7, 8, 10, 11, 12, 16 and 18). For the 9 objects observed 
previously, our observations served to obtain new, independent \HI\ profile 
parameters, as in many cases there were discrepancies between published data
(see Table 2 and Sect. 4.1).
Of the 6 objects in which we clearly detected line emission
(II Zw 96; IC 1623, 2545 and 4395; UGC 9618; NGC 6090 - 
i.e., Nos. 1, 8, 14, 15, 17 and 19), one (IC 2545, No. 8) had not been 
detected before. The \HI\ fluxes measured previously at \nan\ by Martin
et al. (1991) for 4 of these 6 objects are on average about 50\% lower; we prefer 
to use our \nan\ flux values, as they are based on recent external calibrations 
(see Sect. 3).

In all, 8 of the 19 galaxies observed have reliable \HI\ emission line 
detections: the abovementioned 6 we detected, as well as Mrk 463 and UGC 2369 
(i.e., Nos. 3 and 13). Two others show an \HI\ line in absorption: IRAS 03359+1523 and 
10565+2448 (Nos. 4 and 9), and for one (Mrk 848, no. 16) no conclusion
could be drawn due to strong interference.
Global \HI\ profile parameters of (merging) galaxies can depend on the 
telescope used, though, as the different beam sizes (ranging from \am{3}{6} 
diameter at Arecibo to 12$'$ for the Green Bank 90m) may include extended 
gas structures, like tails (e.g., Hibbard \& Yun 1999a), 
or nearby, confusing galaxies. And even observations
made with the same instruments may, surprisingly enough, sometimes show 
significantly different profiles (e.g., for UGC 2369 and II Zw 96). 
See the notes on individual objects in Sect. 4.1. for details.

The derived \HI\ masses listed in Table 3 are plotted as function of 
the projected linear separation between the two merger components in Figure 3.
Concerning the adopted total \HI\ masses for galaxies with more than one 
spectrum available: for the 5 galaxies detected by us we used our \nan\ results,
except for II Zw 96, where we used the average of all 3 available
emission line fluxes; for UGC 2369, we used the Arecibo detection by Haynes et al. (1997)
and for Mrk 463 we used the Arecibo detection by Hutchings et al. (1987) -- 
see also Sect. 4.1.

Six out of the 9 detected objects have a log(\MHI) exceeding 9.8, while 
2 (UGC 2369 and Mrk 463 - i.e., Nos 3 and 13) 
have values $\sim$9.3.
A similar plot for the H$_2$ masses as function of projected separation,
showing a clear correlation, can be found in Gao \& Salomon (1999). 
The relation between the \HI\ mass and separation
will be examined further in Gao et al. (in preparation), using all available
\HI\ line data, both published and others, for a sample of about 30 objects.

%
% -----------------------------------------------------------------
%
% --- Table 3: Global properties
%
\begin{table*}
\bigskip
{\footnotesize
\begin{tabular}{llrrrrrl}
\multicolumn{8}{l}{{\bf Table 3.} Global properties of the sample galaxies}\\
\smallskip \\
\hline
\vspace{-2 mm} \\
  &  & d$_L$ & Sep &  $L_{IR}$ & $L_{CO}$ & log$M_{HI}$ & assumed linewidth \\  % freq. 
  &  & Mpc &  kpc & 10$^{11}$\Lsun\ &  10$^9$$L_{l}$ &  \Msun\ & \kms\ \\       % fact. 
\vspace{-2 mm} \\
\hline
\vspace{-2 mm} \\
1  & IC 1623           &  81.7 &  6.1 &  4.7 & 10.7 &     9.86 &     \\ % 1.017
2  & IRAS 02483+4302   & 213.6 &  3.6 &  6.2 &  2.9 & $<$10.42 & 250 \\ % 1.043 
3  & UGC 2369          & 127.8 & 13.1 &  3.9 &  7.2 &     9.27 &     \\ % 1.026  
4  & IRAS 03359+1523   & 144.5 &  6.5 &  3.3 &  6.9 &  {\it abs.} & 235 \\ % 1.030  
5  & IRAS 04232+1436   & 327.1 &  6.3 & 11.1 &  9.0 & $<$10.08 & 400 \\ % 1.067  
6  & IRAS 08572+3915   & 236.8 &  4.6 & 11.9 &  1.7 & $<$10.35 & 270 \\ % 1.049  
7  & IRAS 10035+4852   & 263.3 &  9.8 &  9.3 &  7.0 & $<$10.60 & 250 \\ % 1.054  
8  & IC 2545           & 137.6 &  3.1 &  4.5 &  2.9 &    10.06 &     \\ % 1.029   
9  & IRAS 10565+2448   & 174.2 &  6.2 &  9.6 &  5.8 &  {\it abs.} &     \\ % 1.037  
10 & Mrk 238           & 204.0 & 13.2 &  2.5 &  6.8 & $<$10.73 & 350 \\ % 1.042  
11 & IRAS 13001-2339   &  86.4 &  2.6 &  2.4 &  2.6 &  $<$9.73 & 450 \\ % 1.018  
12 & NGC 5256          & 110.6 &  4.8 &  3.1 &  5.7 &  $<$9.86 & 400 \\ % 1.023  
13 & Mrk 463           & 204.4 &  3.5 &  5.3 &  2.8 &     9.29 &     \\ % 1.043  
14 & IC 4395           & 147.9 &  4.0 &  2.2 &  4.0 &    10.12 &     \\ % 1.031
15 & UGC 9618          & 137.0 & 25.8 &  4.1 & 17.0 &    10.38 &     \\ % 1.028
16 & Mrk 848           & 162.2 &  4.8 &  7.2 &  7.0 &      --- & 300 \\ % 1.034  
17 & NGC 6090          & 117.6 &  3.5 &  3.0 &  5.0 &    10.15 &     \\ % 1.025
18 & IRAS 20010-2352   & 206.4 &  8.1 &  4.7 &  6.4 & $<$10.96 & 550 \\ % 1.043  
19 & II Zw 96          & 149.9 &  7.4 &  6.6 &  6.0 &    10.09 &     \\ % 1.030
\vspace{-2 mm} \\
\hline
\vspace{-2 mm} \\
\multicolumn{8}{l}{Notes: upper limits to log$M_{HI}$ are 2$\sigma$ values for 
 flat-topped profiles with assumed } \\
\multicolumn{8}{l}{linewidths equal to the measured CO line values; {\it abs} 
 denotes absorption line spectra; } \\
\multicolumn{8}{l}{$L_{l}$ is in units of K km s$^{-1}$ pc$^{-2}$} \\
\end{tabular}
}
\normalsize
\end{table*}
% -----------------------------------------------------------------

\subsection{Notes to individual galaxies}  % 4.1
We searched the vicinity of the target objects for nearby spiral galaxies 
which could possibly give rise to confusion in those \nan\ \HI\ profiles
where line emission was detected. We used the online NED and LEDA databases
(see Webpages http://nedwww.ipac.caltech.edu and 
http://leda.univ-lyon1.fr/pages-html/single.html, repectively), 
in an area of \am{5}{5}$\times$30$'$ ($\alpha\times\delta$) round the pointing 
centre, i.e. about 1.5 times the HPBW, as well as optical images extracted 
from the Digitized Sky Survey (see Fig. 1). For quoted NED values the original 
literature reference is given here (in brackets), while the LEDA data listed are 
weighted average values.

{\bf 1. IC 1623}\, 
Just inside the \nan\ search area, but outside the area shown in Fig. 1, 
lies Sb spiral IC 1622 at an E-W separation of \am{2}{5}, i.e. 1.4 times the 
E-W HPBW, SW of the center of the IC1623A/B pair; its 
$V_{opt}$=6343$\pm$60 \kms, i.e. 285 \kms\ higher than the mean optical velocity 
of the pair, it has magnitude $B_T$=14.53, diameter $D_{25}$=\am{0}{7} 
and an axial ratio of 0.78.
Seen the separation in the sky and in velocity, we do not expect our \nan\
spectrum to be contaminated by this galaxy, unless it has an exceptionally large
\HI\ envelope. Another nearby companion galaxy, nearly 5$'$ NE of 
IC 1623 at only an E-W separation of \am{1}{5}, was detected in the VLA \HI\
maps (for preliminary results, see Hibbard \& Yun 1996; also for detailed maps,
check http://www.cv.nrao.edu/~jhibbard/vv114/), which can only contribute 
less than 10\% of our \nan\ detected \HI\ emission.
Our \nan\ data showed strong baseline curvature in the $H$ polarisation only,
and we therefore used the unaffected $V$ polarisation data for the
data reduction.

{\bf 2. IRAS 02483+4302}\,
In the LEDA database, only a single object (PGC 90441) is listed, while two
are listed in the NED database:  Anon 0248+43A, a 16.52 mag object at 
V=15576$\pm$46 \kms\ (Strauss et al. 1992), and Anon 0248+43B, a 17.36 mag object at 
V=15199 \kms. We used the LEDA entries for PGC 90441 for Table 1.
Our \nan\ data showed stronger baseline curvature in the $V$ polarisation, 
where we therefore fitted a 5$^{th}$ order polyniomial to the data.
The data were affected by strong, narrow radio interference in both
$H$ and $V$ polarisation around 15,525 \kms; this strong, spurious signal has been
blanked out in Figure 1. At full velocity resolution, the narrow absorption 
feature seen at 15,365 \kms\ seems to be due to weaker
interference, as it is much too narrow to be an absorption feature originating
in the merger galaxy.

{\bf 3. UGC 2369}\,        
Within the Arecibo beam (\am{3}{6} HPBW) an \HI\ absorption line
profile (minimum flux density $\sim$-3 mJy) was reported centered at V=9406 \kms\ 
with a $W_{20}$ width of 274 \kms\ by Mirabel \& Sanders (1988), while an 
emission line (of average flux density $\sim$2 mJy) was reported at 
V=9761 \kms\ with a $W_{50}$ width of 229: \kms\ by Haynes et al. (1997), 
who note that the measured width of their profile is poor. The mean optical 
velocities of the 2 galaxies are 9463$\pm$152 and 9380$\pm$118 \kms\ (LEDA), 
respectively, corresponding very well to the velocity of the 
reported absorption line, but not incompatible with the reported 
emission line velocity either, given the uncertainties in the optical velocities.
No line signal was noted in our \nan\ spectrum, with an rms of
2.4 mJy. Mirabel \& Sanders measured a 21 cm continuum flux density 
of 56 mJy.
For the total \HI\ mass, we used the Arecibo detection by Haynes et al. (1997),
which may be a lower limit, however, given the broad CO linewidth  
(FWZI$\sim$900 \kms, Gao \& Solomon (1999)) and possible \HI\
absorption at lower velocity.

% UGC 2369        &  02 51 15.8 &   14 46 14 &  2.43 & <1.22 &       &     &     & N & *     
%                 &             &            &       &  0.48 & 9761 & 229: & --- & A & Ha97  

{\bf 4. IRAS 03359+1523}\, 
In our \nan\ spectrum, with an rms of 1.7 mJy, no line signal was noted, 
neither in emission nor in absorption. Within the Arecibo beam (\am{3}{6} HPBW) 
an \HI\ absorption line profile (minimum flux density $\sim$-5 mJy) was detected by 
Mirabel \& Sanders (1988) and in the present survey. The mean velocity of the
absorption feature, 10721 \kms, is 95 \kms\ higher than the optical systemic
velocity, which has an uncertainty of 45 \kms.
Mirabel \& Sanders measured a 21 cm continuum flux density of 27 mJy.

{\bf 5. IRAS 04232+1436}\, 
Both our \nan\ and Arecibo spectra show a quite narrow absorption 
line (FWHM=90 \kms\ at Arecibo, centered at V=23567 \kms), 
reaching a minimum flux density of about 
5 to 6 mJy. These are likely due to interference, however, as the mean 
velocities of these features are about 270 \kms\ different, and the
measurements were made at 1308 MHz, far outside the protected 1400-1427 
MHz band. The Arecibo spectrum also shows a broader ($W_{20}$=290 \kms) absorption 
feature at the -2 mJy level, too faint for clear detection in \nan, which
may also be due to interference. In conclusion, the data do not
provide sufficient evidence for the detection of an \HI\ line.

{\bf 6. IRAS 08572+3915}\,
In the LEDA database, only a single object (PGC 25295) is listed, while two
are listed in the NED database (IRAS 08572+3915NW and SE), based on near-infrared
imaging observations (Zhou et al. 1993); as positions only are given for these
two components in NED, we have used the LEDA entries for Table 1.

{\bf 7. IRAS 10035+4852}\,
Very close, interacting galaxy pair.
In the LEDA database, only a single object (PGC 29385) is listed, while two
are listed in the NED database: the E and W objects has optical redshifts of,
respectively, 19371$\pm$42 and 19421$\pm$53 \kms (Strauss et al. 1992).
We have used the LEDA entries for Table 1.

% ----- Figure 3 ------------------------------------------------------
%
\begin{figure*}
\vspace*{0.5cm}
% \addtocounter{figure}{-1}
\psfig{figure=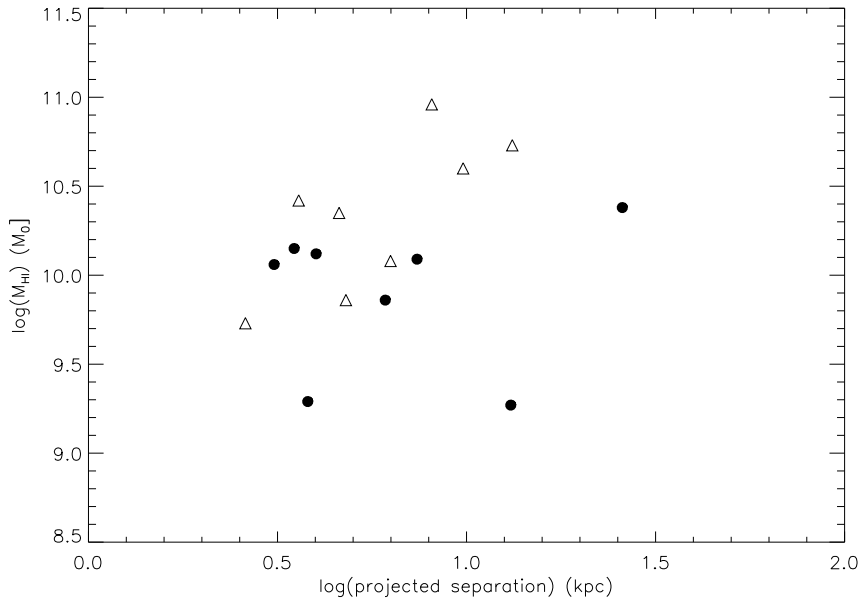,width=12cm}
\caption [] {Total \HI\ gas mass plotted as function of the projected linear
separation between the two components of the 16 mergers observed which
show line emission or for which an upper limit to the line flux could be
estimated; objects which show an absorption line only were excluded.
Filled circles denote detections and triangles upper limits.
Upper limits to the \HI\ masses are 2$\sigma$ values for flat-topped 
profiles with assumed widths equal to the measured CO line width 
of the objects (see Table 3).}
\end{figure*}

{\bf 8. IC 2545}\,       
Very close, interacting galaxy pair.  
In the LEDA database, only a single object (PGC 29334) is listed, while two
are listed in the NED database, based on Lauberts \& Valentijn (1989).
Besides positions, for the NW object a radial velocity (10267$\pm$86 \kms)
and a B magnitude (15.27) are listed, and for the SE object only a B 
magnitude (16.83). We have used the LEDA entries for Table 1.
Detected for the first time in \HI.

{\bf 9. IRAS 10565+2448}\, 
Two distinct, interacting galaxies can be seen on the Digital Sky Survey.
In the LEDA database, only a single object (PGC 33083) is listed, while two
are listed in the NED database; we used the latter for Table 1. 
The optical redshift references are Kim et al. (1995)
for the W object and Strauss et al. (1992) for the NE object.
Within the Arecibo beam (\am{3}{6} HPBW) an \HI\ line profile showing both
emission and much stronger absorption (maximum/minimum flux density $\sim$+1/-5 mJy,
respectively) was reported by Mirabel \& Sanders (1988) - see the two entries in 
Table 2, where the emission line values are lower limits - who measured a 21 cm 
continuum flux density of 44 mJy. 
Our Arecibo observations show some structure in the base line, as they
were taken after sunrise. The deepest and broadest negative feature
in our data corresponds well to the one noted by Mirabel \& Sanders. We
lack the sensitivity, however, to verify their extremely faint ($\sim$-0.5 mJy),
very broad absorption feature extending down to about 12,300 \kms. 
We see no sign of their emission feature adjacent (at higher velocity) to the 
deeper absorption line. 

{\bf 11. \, IRAS 13001-2339 = ESO 507-G070}\, 
Our \nan\ data, obtained over 12 days, showed baseline curvature in both 
$H$ and $V$ polarisation, and we therefore fitted a 5$^{th}$ order
polyniomial to the data.

{\bf 12. NGC 5256}\,  
Both the LEDA and NED databases have entries for 2 objects.
In the LEDA database, a B magnitude is given for one object
only (13.67), while in NED the same B magnitude (14.1) given for
both objects.
      
{\bf 13. Mrk 463}\,   
The Arecibo detection of Mrk 463 (Hutchings et al. 1987) 
at 0.2 Jy \kms\ is at the 4$\sigma$ level, with a peak flux 
density of 1.6 mJy; the authors note that the profile was 
detected in both observing runs.
Heckman et al. (1978) note that their Green Bank  flux measurement 
is relatively uncertain, 6.0$\pm$2.3 Jy \kms, and that it would imply 
a very high \MHILB\ ratio for Mrk 463 if all gas detected resides in 
this galaxy. The reported Green Bank flux is 30 times higher than the 
Arecibo value, and the central velocities of these detections seem 
incompatible: 14702 and 15230 \kms, respectively ($\Delta$V=528 \kms),
for a $W_{20}$ profile width of 200 \kms\ measured at Arecibo.
Inside the Green Bank HPBW but well outside the Arecibo beam,
the DSS shows only one other galaxy that might be a source of
confusion in the Green Bank \HI\ spectrum: CGCG 103-016, \am{6}{1} 
to the SE of the target galaxy, a $B_T$ 15.5 object 
of \am{0}{4} diameter and unknown redshift.
If it were the sole source of \HI\ emission within the Green Bank beam,
it would have an \MHILB\ ratio of 1.4 \MsunLBsun, which is about 
4 times the mean ratio for an Sc spiral (Roberts \& Haynes 1994), but
not unheard of for late-type galaxies.
The morphology of CGCG 103-016 is hard to discern, however, as a relatively
bright star is situated close to its centre. 
In conclusion, CGCG 103-016 may be the cause of the discrepant Green Bank flux.
For the total \HI\ mass, we used the Arecibo detection by Hutchings et al. (1987).

{\bf 14. IC 4395}\,  
Freudling et al. (1988) note that the parameters of their Arecibo \HI\ profile
are uncertain, as it was detected at the edge of the band.
Considerable flux may well have been missed within the Arecibo band,
given the fact that our \nan\ integrated line flux is considerably (factor 4)
larger than the Arecibo flux and that our $W_{50}$ linewidth of
308 \kms\ is 1.5 times larger than the Arecibo value.
Our \nan\ data showed strong baseline curvature in the $V$ polarisation,
and we therefore used only the unaffected $H$ polarisation data for the
data reduction. 
     
{\bf 15. UGC 9618}\, 
All three available spectra (one from Arecibo and 2 from \nan, see Table 2)
show a clear emission line with a pronounced central depression.
Only for the Arecibo spectrum, Mirabel \& Sanders (1988) interpreted this
as an absorption feature, see their two entries in Table 2; they report 
a 21 cm continuum flux density of 109 mJy.
    
{\bf 16. Mrk 848}\,  
The \nan\ data were heavily affected by radio interference, particularly in the 
$H$ but also in the $V$ polarisation, within the range where the \HI\
line emission was expected. Though we used only the least affected data
for the spectrum shown in Fig. 2, the remaining emission features 
appear to be due to RFI.
We therefore do not consider this galaxy as detected in our survey,
but we cannot give a proper upper limit to its integrated line flux,
due to the remnant RFI occuring at its redshift.
    
{\bf 17. NGC 6090}\,     
In the LEDA database, only a single object (PGC 57437) is listed, while two
are listed in the NED database. We used the NED entries for Table 1;
both optical redshifts are from Kim et al. (1995).
  
{\bf 19. II Zw 96}\,        
In the LEDA database, only a single object (PGC 65779) is listed, while four
are listed in the NED database; we used the NED entries of the first two for Table 1.
Near-infrared imaging (Goldader et al. 1997) revealed 4 sources; one (1$^{st}$ in
Table 1) seems to be a relatively quiescent spiral galaxy, another (2$^{nd}$ in
Table 1) may be a galaxy nucleus, since it shows hints of a bar as well as powerful 
starburst activity, and the other two are most likely highly obscured, 
luminous starburst regions. Huge molecular gas concentrations
were discovered recently around these two highly obscured starburst regions 
by Gao et al. (2000a).
All published Arecibo, Green Bank and \nan\ spectra show a clear emission feature. 
Only Mirabel \& Sanders (1988) reported an absorption feature 
(minimum flux density -2.5 mJy) in their Arecibo profile, adjacent to the emission 
line (see Table 2), which does not occur in the Arecibo spectra taken by us or
by Giovanelli \& Haynes (1993); we therefore consider it to be spurious.

\acknowledgements{
We would like to thank Drs. K. O'Neil and T. Ghosh for their help with the 
Arecibo observations and data reduction and  Ms. M. Gendre for her help with 
the Nan\c{c}ay data reduction.
We have made use of the NASA/IPAC Extragalactic Database (NED) which is
operated by the Jet Propulsion Laboratory, California Institute of
Technology, under contract with the U.S. National Aeronautics and Space
Administration, as well as the Lyon-Meudon Extragalactic Database                  
(LEDA) supplied by the LEDA team at the CRAL-Observatoire de                  
Lyon (France).
The Unit\'e Scientifique \nan\ of the Observatoire de Paris is associated as 
Unit\'e de Service et de Recherche USR No. B704 to the French Centre National 
de Recherche Scientifique (CNRS). \nan\ also gratefully acknowledges 
the financial support of the D\'epartement du Cher, 
the European Community, the FNADT and the R\'egion Centre.
}


\begin{thebibliography}{}
\bibitem[]{}
Bieging, J. H., \& Biermann, P. 1983, A\&A, 88, 161
\bibitem[]{}
Bushouse, H. A. 1987, ApJ, 320, 49
\bibitem[]{}
Bushouse, H. A., Werner, M. W., \& Lamb, S. A. 1988, ApJ, 335, 74
\bibitem[]{}
Combes, F., Prugniel, P., Rampazzo, R., \& Sulentic, J. W. 1994, A\&A, 281, 725 
\bibitem[]{}
Condon, J. J., Huang, Z.-P., Yin Q. F., \& Thuan, T. X. 1991, ApJ, 378, 65
\bibitem[]{}
Deich, W. 1990, ANALYZ User's Guide (Arecibo Observatory)
\bibitem[]{}
Downes, D., \& Solomon, P. M. 1998, ApJ, 507, 615.
\bibitem[]{}
Fouqu\'e, P., Bottinelli, L., Durand, N., Gouguenheim, L., \&
 Paturel, G. 1990, A\&AS, 86, 473
\bibitem[]{}
Freudling, W., Haynes, M. P., \& Giovanelli, R. 1988, AJ, 96, 1791 
\bibitem[]{}
Gao, Y. 1996, Ph.D. thesis, State University of New York at Stony Brook 
\bibitem[]{}
Gao, Y., Gruendl, R., Lo, K. Y., Hwang, C. Y., \& Veilleux, S. 1997, 
 in Star Formation: Near and Far, ed. S. S. Holt, \& L.G. Mundy (AIP Press: 
 New York), 319 
\bibitem[]{}
Gao, Y., \& Solomon, P. M. 1999, ApJ, 512, L99
\bibitem[]{}
Gao, Y., Gruendl, R. A., Hwang, C. Y., Lo, K. Y. 1999, in Galaxy Interactions 
 at Low and High Redshift, IAU Symposium 186, ed. J. Barnes, \& D. Sanders 
 (Dordrecht: Kluwer), 227
\bibitem[]{}
Gao, Y., \& Solomon, P. M. 2000a, ApJ, submitted
\bibitem[]{}
Gao, Y., \& Solomon, P. M. 2000b, ApJ, submitted
\bibitem[]{}
Gao, Y., Goldader, J. D., Seaquist, E. R., \& Xu, C. 2000a, in
 Science with the Atacama Large Millimeter Array (ALMA),
 ed. A. Wootten, in press (astro-ph/0008114)
\bibitem[]{}
Gao, Y., Lo, K. Y., Lee, S.-W., \& Lee, T.-H. 2000b, ApJ, in press (astro-ph/0010128)
\bibitem[]{}
 Garwood, R. W., Dickey, J. M., \& Helou, G. 1987, ApJ, 322, 88
\bibitem[]{}
 Genzel, R., Lutz, D., Sturm, E., et al. 1998, ApJ, 498, 569
\bibitem[]{}
Giovanelli, R., \& Haynes, M. P. 1993, AJ, 105, 1271 
\bibitem[]{}
Goldader, J. D., Goldader, D. L., Joseph, R. D., Doyon, R., \& 
 Sanders, D. B. 1997, AJ, 113, 1569
\bibitem[]{}
Haynes, M.P., Giovanelli, R., Herter, T., et al. 1997, AJ, 113, 1197 
\bibitem[]{}
Heckman, T. M., Balick, B., \& Sullivan, W. T. 1978, ApJ, 224, 745
\bibitem[]{}
Hibbard, J. E., \& van Gorkom, J. H. 1996, AJ, 111, 655 
\bibitem[]{}
Hibbard, J. E., \& Yun, M. S. 1996, in Cold Gas at High Redshift,
 ed. M. Bremer et al. (Dordrecht: Kluwer), 47 
\bibitem[]{}
Hibbard, J. E., \& Yun, M. S. 1999a, AJ, 118, 162
\bibitem[]{}
Hibbard, J. E., \& Yun, M. S. 1999b, ApJ, 522, 93 
\bibitem[]{}
Huchtmeier, W. A., \& Richter, O.-G. 1989, A General Catalog of \HI\ 
 Observations of Galaxies (Springer: Heidelberg)
\bibitem[]{}
Hutchings, J. B., Gower, A. C., \& Price, R. 1987, AJ, 93, 6
\bibitem[]{}
 Jackson, J.M., Snell, R. L., Ho, P. T. P., \& Barrett, A. H. 1989, ApJ, 337, 680
\bibitem[]{}
Kennicutt, R. C., Roettiger, K. A., Keel, W. C., van der Hulst, J. M., \& 
 Hummel, E., 1987, AJ, 93, 1011
\bibitem[]{}
 Kennicutt, R. C. 1998, ApJ, 498, 541
\bibitem[]{}
Kim, D.-C., Sanders, D. B., Veilleux, S., Mazzarella, J. M., \& Soifer, B. T. 1995, 
 ApJS, 98, 129
\bibitem[]{}
Lauberts, A., \& Valentijn, E. A. 1989, The Surface Photometry Catalogue of the 
 ESO-Uppsala galaxies (ESO: Garching bei M\"unchen)
\bibitem[]{}
Leech, K. J., Rowan-Robinson, M., Lawrence, A., \& Hughes, J. D. 1994, 
 MNRAS 267, 253 
\bibitem[]{}
Lo, K. Y., Gao, Y., \& Gruendl, R. 1997, ApJ, 475, L103. 
\bibitem[]{}
Martin, J. M., Bottinelli, L., \& Gouguenheim, L. 1991, A\&A, 245, 393 
\bibitem[]{}
Matthews, L. D., van Driel, W., \& Gallagher, J. S. 1998, AJ, 116, 1196
\bibitem[]{}
Matthews, L. D., \& van Driel, W. 2000, A\&AS, 143, 421
\bibitem[]{}
Matthews, L. D., van Driel, W., \& Monnier-Ragaigne, D. 2001, A\&AS, in press
 (astro-ph/0010075)
\bibitem[]{}
Mirabel I. F., \& Sanders, D. B. 1988, ApJ, 335, 104
\bibitem[]{}
Mirabel I. F., \& Sanders, D. B. 1989, ApJ, 340, L53
\bibitem[]{}
Mirabel I. F. 1997, private communication
\bibitem[]{}
Murphy, T. W., Armus, L., Matthews, K., et al. 1996, AJ, 111, 1025 
\bibitem[]{}
Roberts, M. S., \& Haynes, M. P. 1994, ARA\&A, 32, 115
\bibitem[]{}
 Sanders, D. B., Soifer, B. T., Elias, J. H., et al. 1988, ApJ, 325, 74 
\bibitem[]{}
Sanders, D. B. 1992, in ASP Conf. Ser. 31, Relationships between Active 
 Galactic Nuclei and Starburst Galaxies, ed. A. Fillipenko, 303 
\bibitem[]{}
Sanders, D. B., \& Mirabel I. F. 1996, ARA\&A, 34, 749 
\bibitem[]{}
Sanders, D. B. 1999, Ap\&SS 266, 331 
\bibitem[]{}
Scoville, N. Z., Sargent, A. I., Sanders, D. B., \& Soifer, B. T. 1991, 
 ApJ, 366, L5
\bibitem[]{}
Scoville, N. Z., Evans, A. S., Thompson, R., et al. 2000, AJ, 119, 991 
\bibitem[]{}
Smith, H. E., Lonsdale, C. J., \& Lonsdale, C. J. 1998, ApJ, 492, 137
\bibitem[]{}
Soifer, B. T., Neugebauer, G., Matthews, K., et al. 2000, AJ, 119, 509 
\bibitem[]{}
 Solomon, P. M., Downes, D., \& Radford, S. J. E. 1992, ApJ, 387, L55 
\bibitem[]{}
Solomon, P. M., Downes, D., Radford, S. J. E., \& Barrett, J. 1997, 
 ApJ, 478, 144
\bibitem[]{}
Strauss, M. A., Huchra, J. P., Davis, M., et al. 1992, ApJS, 83, 29
\bibitem[]{}
Toomre, A., \& Toomre, J. 1972, ApJ, 178, 623.
\bibitem[]{}
Toomre, A. 1977, in The Evolution of Galaxies and Stellar Populations, 
 ed. B.M. Tinsley B.M., \& R.B. Larson (New Haven: Yale Univ.), 401
\bibitem[]{}
van den Broek, A. C., van Driel, W., de Jong, T., et al. 1991, A\&AS, 91, 61 
\bibitem[]{}
Veilleux, S., Sanders, D. B., \& Kim, D.-C. 1997, ApJ, 484, 92
\bibitem[]{}
Veilleux, S., Sanders, D. B., \& Kim, D.-C. 1999, ApJ, 522, 139
\bibitem[]{}
Young, J. S., Schloerb, F. P., Kenney, J. D., \& Lord, S. D. 1986, ApJ, 304, 443
\bibitem[]{}
Zhou, S., Wynn-Williams, C. G., \& Sanders, D. B. 1993, ApJ, 409, 149
%
\end{thebibliography}
\end{document}